\documentclass[twocolumn,twoside,slac]{revtex4}
\usepackage{graphicx}
\usepackage{fancyhdr}
\renewcommand{\sc}   {\scshape}
\renewcommand{\tt}   {\ttfamily}

\begin{document}

\title{Simulation Application for the LHCb Experiment}

\author{I. Belyaev$^2$, Ph. Charpentier$^1$, S. Easo$^3$,  
P. Mato$^1$, J. Palacios$^4$, W. Pokorski$^1$, F. Ranjard$^1$ and
 J. van Tilburg$^5$
}
\affiliation{$^1$European Laboratory for Particle Physics (CERN), Gen\`eve,
  Switzerland \\ 
$^2$Institute for Theoretical and Experimental Physics (ITEP), Moscow,
  Russia \\ $^3$Rutherford Appleton Laboratory (RAL), Chilton, UK \\
  $^4$University of Liverpool, Liverpool, UK \\ $^5$National Institute for Nuclear 
  Physics and High Energy Physics (NIKHEF), Amsterdam, Netherlands} 

\begin{abstract}
We describe the LHCb detector simulation application ({\sc{Gauss}}) based on the 
{\sc{Geant4}} toolkit. 
The application is built using the {\sc{Gaudi}} software framework, which is used for all 
event-processing applications in the LHCb experiment. The existence of an underlying 
framework allows several common basic services such as persistency, interactivity, 
as well as detector geometry description or particle data to be shared between simulation,
reconstruction and analysis applications. The main benefits of such common services are 
coherence between different event-processing stages as well as reduced development effort.
The interfacing to {\sc{Geant4}} toolkit is realized through a fa\c{c}ade ({\sc{GiGa}}) which minimizes 
the coupling to the simulation engine and provides a set of abstract interfaces for 
configuration and event-by-event communication. 
The {\sc{Gauss}} application is composed of three main blocks, i.e. event
generation, detector response simulation and digitization which reflect the different 
stages performed during the simulation job. 
We describe the overall design as well as the details of {\sc{Gauss}}
application with a special emphasis on the configuration and control of the underlying 
simulation engine. We also briefly mention the validation strategy and the planing for the LHCb 
experiment simulation.
\end{abstract}

\maketitle
\thispagestyle{fancy}

\section{Introduction}
The LHCb \cite{lhcb} experiment is one of the four main experiments that will be performed using
the Large Hadron Collider (LHC), presently under construction at the European Organization
for Nuclear Research (CERN) in Geneva, Switzerland. The
experiment is devoted to the precision measurements of CP violation in the B
meson system. 

The simulation applications are of
major importance both for the studies during the construction phase as well
as during the running of the detector. The assumption is that they  
produce data (``digits'') that would normally be
available from the electronics of the detector. This simulated data is then fed,
in the same way as the real data would be, 
into the reconstruction and further on into the analysis applications.  The LHCb
collaboration is moving to a complete chain of object-oriented event processing
software. With the new reconstruction and analysis applications already used in
 production, the simulation one is getting close to that stage as well.

The new object-oriented LHCb detector simulation application ({\sc{Gauss}}) is based on the 
{\sc{Geant4}} \cite{g4} toolkit and is currently entering its final
phase of development. It is built within the LHCb software framework
{\sc{(Gaudi)}} \cite{gaudi} which is also used by the rest of the LHCb event-processing
software. The existence of the common underlying framework allows sharing of
many basic services such as persistency, interactivity, histograming, etc. In
addition, it provides  
a natural way of having common sources of data such as the detector geometry,
the magnetic field map, or the particle properties. The actual interfacing of {\sc{Geant4}} toolkit
to the Gaudi framework is done by {\sc{GiGa}} (Gaudi Interface to {\sc{Geant4}} Application)
 \cite{giga}.

The organization of this paper is the following. We will start with a brief
presentation of the general structure of the Gaudi framework followed by a
description of the structure of  
{\sc Gauss} application itself. After that, we  will discuss a few
selected aspects of {\sc{GiGa}} which are particularly relevant as the functional
design of {\sc{Gauss}} is concerned. Finally we will briefly present a few results of
our physics validation and we will formulate some conclusions.

\section{Overview of the underlying software framework}
The philosophy of the Gaudi framework is
that each application consists of a set of ``algorithms'' which represent some
self-contained functional parts of the programme. Those algorithms behave like
pluggable components. They can be easily added, removed, executed in series or
separately. The job configuration, i.e. specification of the sequence of algorithms to be
executed, setting of their options, defining input and output files, etc, is done via job
options files, which are simply text files, read at the beginning of each job
execution. 

On the other hand, the existence of the so-called transient data stores which are used for the 
communication between the algorithms, guarantees the lack of mutual dependence between them. 
For instance, if an algorithm uses as an input MCHits to produce
MCDigits, the only thing it requires is that there are some MCHits available in
the transient store. It does not care whether those MCHits come from a simulation
algorithm which was executed beforehead or from a data file which was used to populate the
transient store. Moreover, if at the moment when the algorithm requested the
MCHits from the transient store they were not there yet, it will be the
framework who will take care of reading the appropriate file (if available of
course) in order to populate the store. From the point of view of the algorithm,
it will be a completely transparent procedure, and in particular the
implementation of the algorithm will not depend in any way on the persistent
technology used.

Such an architecture allows a modular structure of the whole simulation
programme, introducing less coupling between different stages of the
simulation job. The existence of the transient stores,
provides also natural ``communication channels'' with other sets of algorithms such as the
reconstruction algorithms or the visualization algorithms. In some sense, an
``application'' becomes a less well-defined entity, since one is able to combine
any collection of algorithms into one executable sequence. One can imagine, for
instance, running the simulation algorithms together with the visualization
ones, providing a display of each simulated event. It is useful to stress
here once again, that those different algorithms can be developed
completely independently without implementing any specific interfaces between
them (apart from the ones defined by the framework to communicate with the
transient stores).

\section{Overview of the simulation application}
\begin{figure}
\includegraphics[width=65mm] {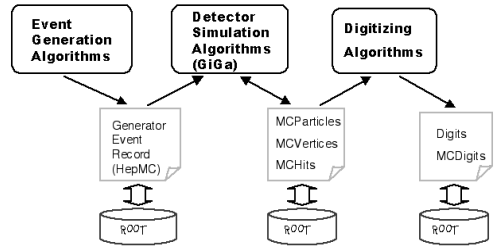}
\caption{Structure of the {\sc{Gauss}} simulation application}
\label{figgauss}
\end{figure}

The {\sc{Gauss}} simulation application consists of three major blocks responsible for
the event generation, detector response simulation and digitization (see
Figure~\ref{figgauss}). These three 
blocks are completely independent and, using the features of the underlying
framework described in the previous section, can be executed   
separately reading (saving) the data from (to) the persistent stores. The
format used to store the generated event (the output from the MC generators) is
the HepMC event record \cite{hepmc}. As far as the simulated event (the output from the
detector simulation) is concerned, {\sc{Gauss}} uses the LHCb event format
\cite{event}, which is then 
used as the input format for the reconstruction applications. As mentioned
before, due to the existence of the underlying software framework, providing in
particular the persistency service, the simulation application is fully
independent from the underlying persistent technology. At the present moment,
for instance, we are using the ROOT \cite{root} file format to save the information on the
disk, but this could be changed at any time, without affecting the actual
simulation application. 

We will now go a little bit more into details and describe the first two blocks
from Figure~\ref{figgauss}, i.e. the event generation and the interfacing to the simulation
toolkit. The third block, being the digitization, is very detector-specific and
therefore we will not discuss it here.

\subsection{Event generation}
The main event generator for {\sc{Gauss}} is {\sc{Pythia}} \cite{pythia}. It is
used to produce both the minimum-bias as well as the signal events. The actual
decays of the b-hadrons are performed, however, using a dedicated decay package called
{\sc{EvtGen}} \cite{evtgen}. Both {\sc{Pythia}}, as well as {\sc{EvtGen}} are
wrapped inside special  
Gaudi algorithms (called PythiaAlg and EvtDecayAlg, respectively) which make
them to behave like pluggable components (callable and controllable) in the Gaudi
framework. The  
configuration of those MC generators can, therefore, be done via the Gaudi job options
interface.

\begin{figure}
\includegraphics[width=65mm] {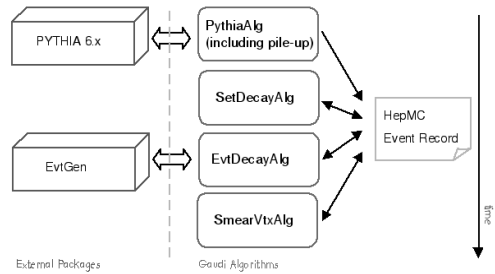}
\caption{Event generation sequence}
\label{figgener}
\end{figure}

As mentioned before, the generated events are stored in the HepMC event record,
which also serves to communicate between different generators (decay
packages). The entire sequence of the event generation is shown on Figure~\ref{figgener}. First
{\sc{Pythia}} is called to generate the primary event. The settings of
{\sc{Pythia}} are such that most of the physical particles (those known to
{\sc{Geant4}}) are declared stable, and therefore the generated event consists,
at this stage, of one physical vertex (store in the transient store in the HepMC
format). Such an approach seems to be natural from the
point of view of the detector response simulation which is performed
afterwards. The exception, however, are the decays of the b-hadrons. Those
are the most important in our simulation, and we want to have full control over
them in order to be able to explicitly generate particular decay channels. 
In order to do so, the selection algorithm (SetDecayAlg) is called, which marks
particles (normally b-hadrons) that should be decayed by the specialized
package (the others will be left to be decayed by {\sc{Geant4}}).  
Once that selection is performed, the EvtDecayAlg is executed. It calls
{\sc{EvtGen}} for each particle marked beforehead and adds the decay products to
the HepMC tree. {\sc{Geant4}} will later on use that information in the context
of the forced decay channel mechanism. 

The final step in the event generation sequence is to call an additional
algorithm which simulates the smearing of the position 
of the primary vertex (due to the size of the colliding beams). It generates the
spread in the x, y and z directions according to distributions specific for
the LHC beams, which are then used to shift the position of all the vertices in
the event. 

Such an architecture is very flexible and can be easily extended. It is very
easy to add other (replace) event generators, without any change to the code of
the remaining simulation chain. Having more than one event generator implemented 
(wrapped) in the form of the Gaudi algorithm, one can run different  
configurations (replacing, for example {\sc{Pythia}} with some other primary
generator) without any recompilation but simply with different job options.  

\subsection{Interface to the simulation toolkit}
The detailed description of the technical aspects of the Gaudi Interface to
{\sc{Geant4}} Application ({\sc{GiGa}}) can be found in \cite{giga} and
therefore we will not repeat it here. We will rather concentrate on a few
selected aspects of {\sc{GiGa}} which 
are particularly interesting from the point of view of the overall simulation
application. 

\begin{figure}
\includegraphics[width=65mm] {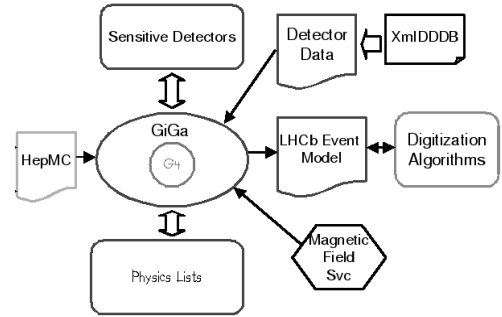}
\caption{Interface between {\sc{Geant4}} and {\sc{Gaudi}} environment}
\label{figgiga}
\end{figure}

As mentioned before, one of the advantages of building the simulation
application on top of the common software framework is that it allows us to use
sources of detector data, i.e. geometry description, magnetic field map, etc,
shared with other applications such as the reconstruction or the
visualization. The 
conversion of the LHCb geometry model to the {\sc{Geant4}} geometry tree is done
at the beginning of each simulation job by a specialized geometry conversion
service.  It is 
worth noting that the actual geometry configuration can be also changed 
via the job options, without any need for recompilation.  The specification of
``sensitive volumes'' is done via simple declarations of the corresponding
sensitive detectors class names in the persistent description
of the shared geometry data (for which we use the XML format). One can add, remove or
change sensitive detectors by simply editing the geometry description
file. Those changes are then taken into account at the runtime and the
instantiation of the required sensitive detectors is done using the abstract
factory approach. The overall structure of the way  
{\sc{Geant4}} is interfaced to the {\sc{Gaudi}} environment is shown in 
Figure~\ref{figgiga}. 

As we can see there, the actual {\sc{Geant4}} engine is put behind a fa\c{c}ade
pattern 
which provides a unified high level abstract interface. A set of abstract
interfaces implemented within {\sc{GiGa}} allows loading from the {\sc{Gaudi}}
transient stores into {\sc{Geant4}},
detector geometry, primary event data, etc.
It also makes it possible to use standard {\sc{Gaudi}} services such as
Magnetic Field service or Job Options service to be used by the {\sc{Geant4}}
application. Let us mention as well that {\sc GiGa}, through the existence of a
specialized class called {\tt GiGaRunManager}, provides internal access to   
the {\sc{Geant4}} event loop. The standard {\tt G4RunManager::BeamOn} method is
never used in the context of {\sc GiGa}. The construction of the primary events is
done via direct calls to {\tt G4Event::AddPrimaryVertex} method and 
the simulation is triggered by direct calls to the {\tt
G4EventManager::ProcessOneEvent} method. This gives us more control
over the each simulation run and provides us an extra flexibility as far as the
construction of the primary events is concerned.

Probably the most interesting feature of {\sc GiGa} as the interface to the
underlying framework is that all the ``actions''
such as event action, tracking action, etc, are instantiated using the abstract
factory approach and therefore can be loaded at the runtime. Such a design
insures a very flexible structure of the simulation program and allows running
different configuration by changing only the job options. As far as the physics
lists are concerned, they are implemented in the form of {\sc{Geant4}} modular
physics list and also are instantiated via the abstract factories. This again
makes the architecture flexible, facilitating validation of different physics
lists without a need of recompilation.

\begin{figure}
\includegraphics[width=65mm] {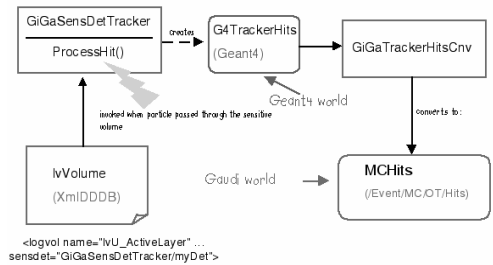}
\caption{Sensitive detectors and hits in Gauss}
\label{sensdet}
\end{figure}

Let us illustrate the interplay between
the underlying framework and the {\sc Geant4} toolkit on the concrete example of
the sensitive detectors and hit creation. Figure~\ref{sensdet} shows the whole
chain starting from the sensitive volume declaration and ending with the
creation of the LHCb event model hits. 

Once {\sc Geant4} is aware of the given sensitive detector (instantiated using
an abstract factory, as described at the beginning of this section), it will
follow the standard procedure of calling the {\tt ProcessHit} method every time
there is a particle passing by the volume associate to it. The
possible result of that (if the required conditions are satisfied, like the
non-zero charge of the passing particle, etc.) will be the
creation of {\sc Geant4} hits, stored in the {\sc Geant4} hit collections. The
last part of the chain will be the conversion of all the created 
{\sc Geant4} hits during the given event, into the LHCb event model hits stored
afterwards 
in the {\sc{Gaudi}} transient data store (and available for the other algorithms
like digitization, or for the serialization). Such a conversion is performed by a
specialized {\sc GiGa} converted at the end of each event. 

From this example we can see that the user doesn't actually deal directly with
the ``{\sc Geant4} world''. He declares the sensitive volumes in the ``{\sc Gaudi}
 world'' and he gets the final output (simulated hits and particles) back in the
{\sc Gaudi} transient store.  

\section{Physics validation}

The physics validation of the new LHCb simulation program is done in two ways:
by comparing with the test beam data and by comparing with the {\sc{Geant3}}
simulation results. The former approach is used to validate particular processes
like the Rayleigh scattering or the Cherenkov radiation in the material of the Ring
Imaging Cherenkov counters (RICH), or
hadronic interaction models for the calorimeter. The later approach is used
more to test the overall simulation chain, together with the event generation
phase. What we will briefly present in this section is a few comparison plots
between the {\sc Geant3} and {\sc Geant4} results for the tracker-like devices
in the LHCb.

\subsection{Vertex Locator physics validation}

The Vertex Locator (VELO) is the LHCb subdetector inside which the
interaction point is placed. It consists of a series of Silicon sensors and its
purpose is to perform precise measurements of tracks close to the interaction
region. Those measurements are essential for the reconstruction of production and decay
vertices of the beauty- and charm-hadrons.

\begin{figure}
\includegraphics[width=65mm] {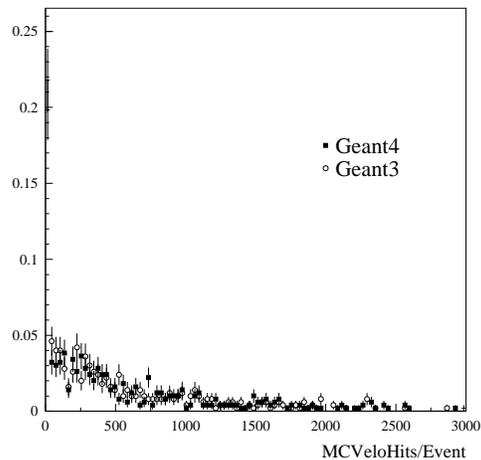}
\caption{Number of simulated VELO hits per event}
\label{velomult}
\end{figure}

\begin{figure}
\includegraphics[width=65mm] {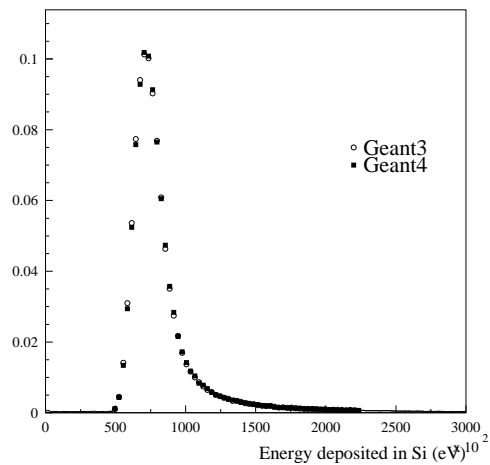}
\caption{Energy deposition in the VELO Silicon sensors}
\label{veloenergy}
\end{figure}

We have examined several different histograms (multiplicities, distributions in
space, energy depositions, time of flight, etc) produced using simulated data
and they all show a good agreement between the {\sc Geant3} and {\sc Geant4} based 
applications. In the Figures~\ref{velomult} and \ref{veloenergy} we
see for instance the multiplicity of simulated VELO hits and the energy
deposition in the Silicon of sensors. 

\begin{figure}
\includegraphics[width=65mm] {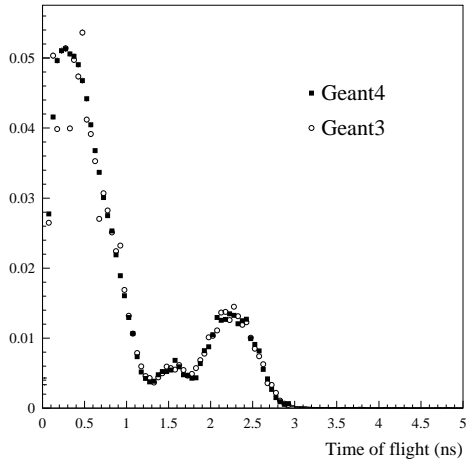}
\caption{Time of flight distribution for particles creating hits in the  VELO
  Silicon sensors} 
\label{tofvel}
\end{figure}

In some cases, like for example for the time of flight distribution shown on
Figure~\ref{tofvel}, we see that {\sc Geant4} plot looks actually ``smoother''
for the lower values, which is probably due to the higher precision compared to
{\sc Geant3}.

\subsection{Outer Tracker physics validation}
The outer tracking detector in LHCb consist of drift cells with straw-tube
geometry. In the configuration of the tracking stations the emphasis is on
tracking precision in the (x,z) magnet bending plane.  

\begin{figure}
\includegraphics[width=65mm] {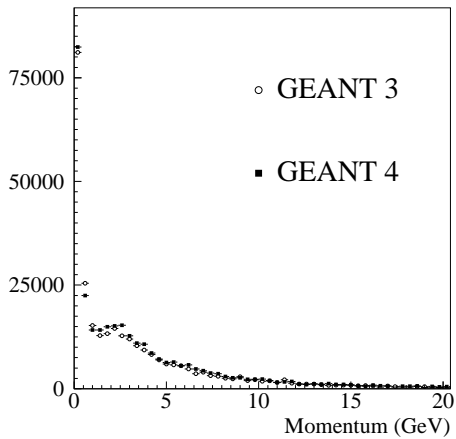}
\caption{Momentum distribution for particles creating hits in the Outer Tracker}
\label{otmomentum}
\end{figure}

\begin{figure}
\includegraphics[width=65mm] {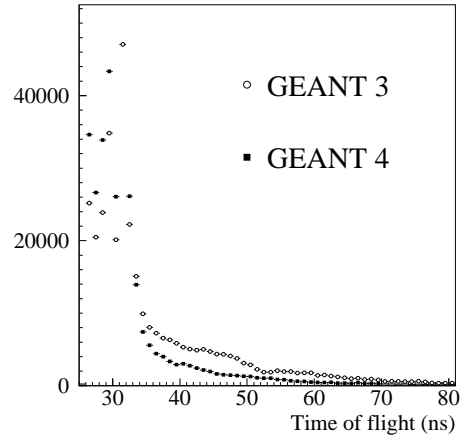}
\caption{Time of flight distribution for particles creating hits in the Outer Tracker}
\label{ottof}
\end{figure}

In order to validate the new simulation program, we have again compared a number 
of different distributions. In most of the cases, as we can see for instance for 
the momentum distribution of particles creating hits shown in
Figure~\ref{otmomentum}, we have not observed any
significant discrepancies between the events simulated by {\sc Geant4} and {\sc
 Geant3}. The only clearly  visible difference was in the time of flight
distribution, which we can see in Figure~\ref{ottof}.

It seems that in the {\sc Geant3} simulation there are more particles with
longer time of flights (and less with shorter ones, since the overall
multiplicity seems to be the same) than in the {\sc Geant4}
simulation. This result is not yet understood and we are currently investigating
any possible cause of that difference. 

\section{Status and conclusions}

Over the last year the {\sc Geant4} based simulation application for the LHCb
experiment has evolved from a set individual components to a fully functional
program. Its major parts such as the interfaces to Monte-Carlo
generators, the interface to the {\sc{Geant4}} toolkit, as well as
different sensitive detectors are, to a large extend, implemented. Due to the
underlying software framework, many of the components, such 
as the detector description, are shared between the simulation as well as the
other applications like the reconstruction or the visualization. The
interface between {\sc Geant4} and that framework extensively uses the concept
of abstract factories which makes the simulation environment very flexible and
easily configurable.  

The LHCb is now testing its new simulation application. With most of the results
being positive, the move to the {\sc Geant4} based simulation is being
prepared. The LHCb is planing to start extensive test productions in the summer
2003 and to move definitely to the {\sc Geant4}-based simulation application at
the beginning of the year 2004.

\end{document}